\documentclass[11pt]{article}

\usepackage{comment}
\usepackage{algorithm}
\usepackage{algpseudocode}
\usepackage{amsmath}
\usepackage{graphicx,psfrag}
\usepackage{graphicx} 
\usepackage{enumerate}
\usepackage{url} 
\usepackage{mathptmx}
\usepackage{amsthm}
\usepackage{amssymb}
\usepackage{mathtools}
\usepackage{appendix}
\usepackage{here}
\usepackage[T1]{fontenc}
\usepackage{color}
\usepackage{xcolor}
\usepackage{setspace}
\usepackage[utf8]{inputenc}
\usepackage{subfig}
\usepackage[normalem]{ulem}

\newcommand{\R}{{\mathbb R}}

\def\E{{\mathbb E}}

\newcommand{\BB}{{\mathcal B}}
\newcommand{\1}{{\mathbf 1}}

\newcommand{\bea}{\begin{eqnarray}}
\newcommand{\eea}{\end{eqnarray}}
\newcommand{\be}{\begin{equation}}
\newcommand{\ee}{\end{equation}}
\newcommand{\beann}{\begin{eqnarray*}}
\newcommand{\eeann}{\end{eqnarray*}}
\newcommand{\balnn}{\begin{align*}}
\newcommand{\ealnn}{\end{align*}}

\title{Statistical modeling of diabetic neuropathy:  {\sc E}xploring the dynamics of nerve mortality}

\author{Konstantinos Konstantinou$^1$, Farnaz Ghorbanpour$^2$, Umberto Picchini$^1$, \\ Adam Loavenbruck$^3$ and Aila S\"arkk\"a$^1$ }

\date{%
    $^1$Department of Mathematical Sciences, Chalmers University of Technology and the University of
Gothenburg, Gothenburg, Sweden.\\%
   $^2$Department of
Mathematical Sciences, Allameh Tabataba'i
University, Tehran,  Iran.\\%
  $^3$Department of Neurology, Kennedy Laboratory, University of Minnesota,
Minneapolis, Minnesota\\[2ex]%
}
\begin{document}

\maketitle

\thispagestyle{plain}			
\setlength{\parskip}{0pt plus 1.0pt}
\section*{Abstract}
Diabetic neuropathy is a disorder characterized by impaired nerve function and reduction of the number of epidermal nerve fibers per epidermal surface. Additionally, as neuropathy related nerve fiber loss and regrowth progresses over time, the two-dimensional spatial arrangement of the nerves becomes more clustered. These observations suggest that with development of neuropathy, the spatial pattern of diminished skin innervation is defined by a thinning process which remains incompletely characterized. We regard samples obtained from healthy controls and subjects suffering from diabetic neuropathy as realisations of planar point processes consisting of nerve entry points and nerve endings, and propose point process models based on spatial thinning to describe the change as neuropathy advances. Initially, the hypothesis that the nerve removal occurs completely at random is tested using independent random thinning of healthy patterns. Then, a dependent parametric thinning model that favors the removal of isolated nerve trees is proposed. Approximate Bayesian computation is used to infer the distribution of the model parameters, and the goodness-of-fit of the models is evaluated using both non-spatial and spatial summary statistics. Our findings suggest that the nerve mortality process changes behaviour as neuropathy advances. \\


\noindent Keywords: 
Approximate Bayesian computation; Dependent thinning; Epidermal nerve fibers; Random thinning; Reactive territory; Spatial point process  

\section{Introduction}

Epidermal nerve fibers (ENFs) are dendroidal thin sensory nerve fibers in the outermost layer of the human skin, called epidermis. They enter, grow and branch in the epidermis until they terminate. Throughout the paper, the entry points will be referred to as base points and the termination points as end points. The nerve fibers transfer signals such as heat and pain recorded by the end points to the central nervous system. Diabetic neuropathy is a disorder that in which elevated blood sugar and related processes in the body damage ENFs all along their course, from the dorsal root ganglion near the spinal cord to the skin, negatively affects their functionality, and over time causes attrition of ENFs. Progression of ENF dysfunction and loss is characterized, respectively, by neuropathic  pain and loss of sensation \cite{kennedy1996quantitation}. While damaged nerves may heal and regrow with sustained improvement of blood sugar, this regrowth is very slow and often incomplete. It is therefore important to detect the neuropathy at the earliest stage possible, to prevent ENF damage before it occurs.

The diagnostic capabilities of the ENFs have been established in several studies. More specifically, neuropathy progression decreases the ENFs spatial intensity and total coverage of the ENFs in the epidermis \cite{kennedy1996quantitation,andersson2016discovering,ghorbanpour2021marked}. In addition,  the two-dimensional spatial structure of the base and end points of subjects suffering from diabetic neuropathy tend to be more clustered than the structure of healthy controls \cite{waller2011second,myllymaki2012analysis,andersson2016discovering,olsbo2013development}.
Furthermore, the nerve fibers in subjects with diabetic neuropathy tend to branch fewer times before terminating than the nerve fibers in healthy subjects \cite{andersson2019bayesian}.

 A considerable amount of earlier research on the spatial structure of ENFs has concentrated on modelling the spatial structure. The planar locations of the base and end points are treated as realisations of two-dimensional spatial point processes, and point process models have been developed for the end points conditioned on the empirical base point locations. For instance, the non-orphan cluster (NOC) model \cite{olsbo2013development} and the uniform cluster centre (UCC) model \cite{andersson2016discovering} are point process models of this nature. In the UCC model, the direction of the end point clusters with respect to their corresponding base points is uniformly distributed, while in the NOC model the clusters are constructed towards open space.  To capture possible interactions between the entire nerve trees, a sequential marked point process model was proposed in Ghorbanbour et al.\ \cite{ghorbanpour2021marked}. Furthermore, a continuous time birth-and-death process that allows interactions between the base points and within the points in each end point cluster was proposed in Garcia et al. \cite{garcia2020interacting}. In addition, some models for the three-dimensional spatial structure have recently been suggested in \cite{konstantinouspatial,konstantinou2022pairwise}

In this paper, we will focus on the underlying  process that guides the morphological changes in the spatial structure of the nerve trees as diabetic neuropathy advances. We have skin samples from healthy subjects and subjects suffering from either mild or moderate diabetic neuropathy. The mild  point patterns consisting of the base and end points of ENFs are treated as spatial thinnings of the healthy point patterns, and the moderate patterns as thinnings of the mild patterns. For this purpose, different spatial thinning schemes are proposed. Since for such thinning models we do not have a likelihood function readily available, we suggest an approximate Bayesian computation (ABC) approach to estimate the parameters of the model.  Finally, the models are evaluated using  Ripley's $K$ function, mark correlation function, and some non-spatial summary statistics. Our findings indicate that nerve mortality does not occur completely at random.  To the best of our knowledge, this is the first study investigating nerve loss due to  diabetic neuropathy using spatial thinning models. 

The paper is organised as follows. In Section 2, the ENF data set is described and a brief introduction to point processes and spatial thinning operations is given in Section 3. In Section 4, the proposed thinning schemes are described. Our findings are presented in Section 5 and further discussed in Section 6. 

\section{Data}
 The epidermal nerve fiber data we have available are obtained using suction induced skin biopsies, a medical procedure where a skin sample is taken, mounted on a slide and stained for imaging \cite{wendelschafer2005epidermal,panoutsopoulou2009skin}. Then, confocal microscopy is used to manually trace the base points, which are the entry locations of the ENFs in the epidermis, the branching points,which are the locations where the nerve branches within epidermis, and the end points, which are the locations where the nerve fibers terminate.  Two skin blister specimens were taken from different body parts from each subject in the study resulting in three to six images (usually four) per subject and body part. The degree of diabetic neuropathy, i.e.\ healthy, mild, moderate, or severe, is known for each subject.  The original spatial point patterns are three dimensional. However, since we are interested in the coverage of the ENFs on the skin, we concentrate on the two dimensional projections of the patterns.

\begin{figure}[H]
    \centering\includegraphics{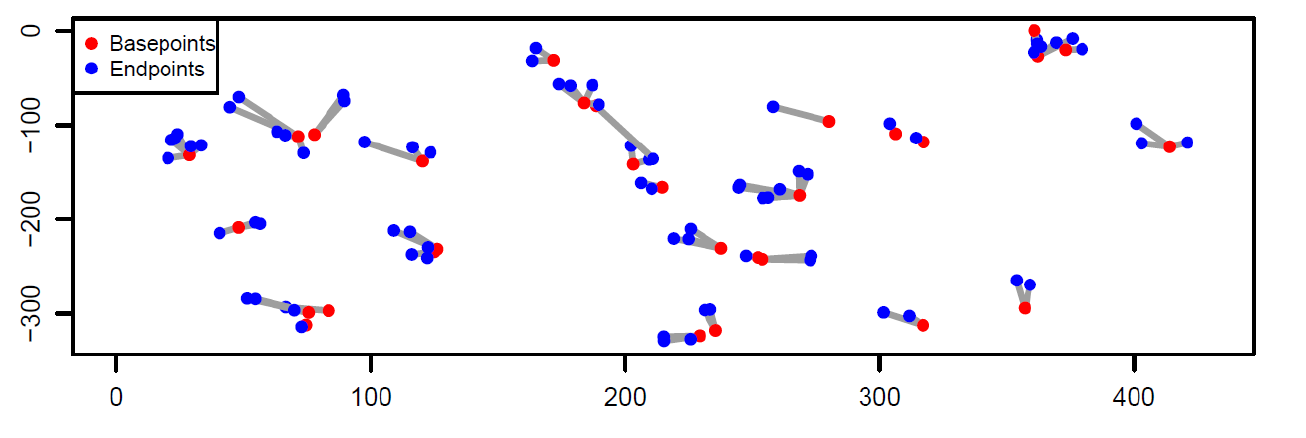}
    \caption{An illustration of the structure of the nerve trees in a healthy sample. Red points represent the base and blue points the end points of the nerve fibres. }
    \label{fig:data}
\end{figure}

 Here, we limit our analysis to the data collected from the feet of 32 healthy subjects,  8 subjects with mild diabetic neuropathy, and 5 subjects with moderate diabetic neuropathy.  The choice of the body part is motivated by the observation that changes in the ENFs morphology occur at the earliest stage in distant body parts such as feet \cite{kennedy1999}. 
 We have left out the group with severe diabetic neuropathy as those samples  contain very few nerves. The data consisting of the base and end point locations are treated as realisations of spatial point processes in $\mathbb{R}^2$ observed in the  window $W$ with area of approximately 330 $\times$ 432 microns. In total, 112 healthy, 28 mild diabetic and 13 moderate diabetic skin samples are included  in the analysis. From now on, we refer to the three groups as healthy, mild, and moderate, respectively. An example of an ENF sample is displayed in Figure \ref{fig:data}, where the different types of points are represented with different colours. Examples for ENF samples from mild and moderate diabetic samples are shown in Figure \ref{fig:samples_ex} in Appendix.

\begin{figure}[H]
\centering\includegraphics[scale=1]{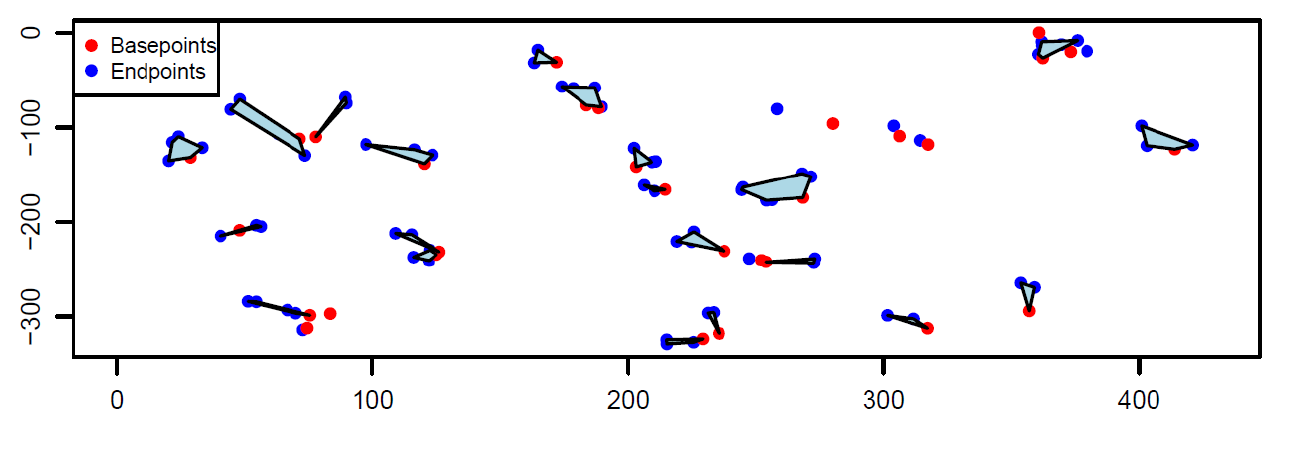}    \caption{Reactive territory of a skin sample obtained from the same  healthy subject as in Figure \ref{fig:data}. Red points represent the base and blue points the end points of the nerve fibres.}    \label{fig:reactive}
\end{figure}

The area of the skin that the ENFs cover can be described by  reactive territories introduced in Andersson et al.\ \cite{andersson2016discovering}. The reactive territory of a nerve tree is defined as the convex hull determined by the locations of the projected end points and base points belonging to the same nerve tree. An example of a reactive territory for  a healthy sample is presented in Figure \ref{fig:reactive}. Note that a nerve tree has to have at least two end points to have a positive reactive territory.

\begin{figure}[H]
    \centering\includegraphics[scale=0.6]{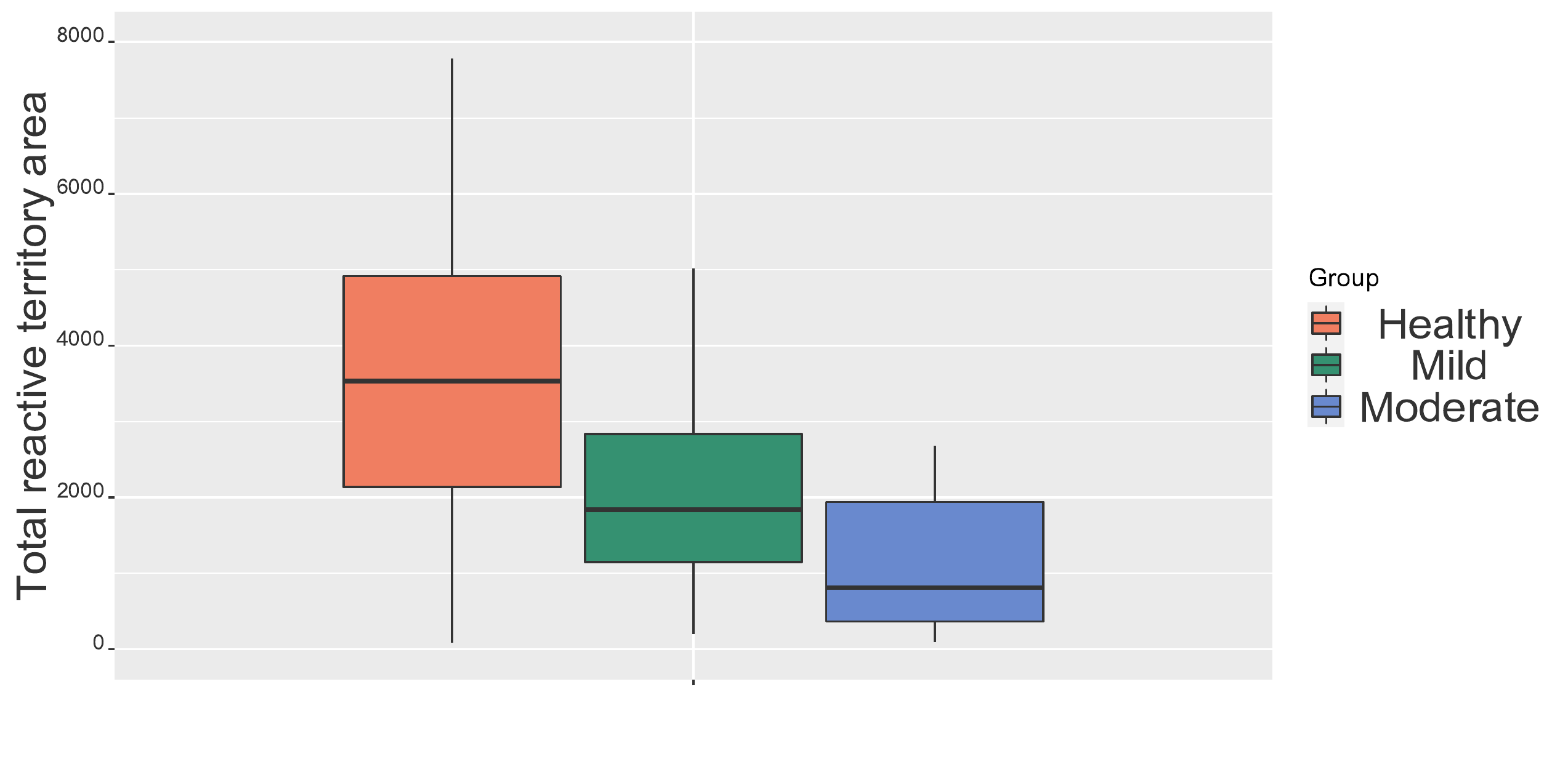}
    \caption{The total area of the skin covered by the reactive territories in the samples from the healthy subjects and the subjects with mild or  moderate diabetic neuropathy. }
    \label{fig:reactives}
\end{figure}

It is well established that the degree of neuropathy and the area of positive reactive territory are negatively correlated, i.e.\ as the degree of neuropathy increases the area of the skin covered by the ENFs decreases \cite{kennedy1996quantitation,andersson2016discovering}. This is illustrated for our data in Figure \ref{fig:reactives}. The decrease in ENF coverage is translated into neuropathic pain and loss of sensation, the main symptoms of the neuropathy.

\section{Methods for spatial point processes  }
 Point patterns consisting of  ENF base and end points projected into the plane are regarded as realisation of  two dimensional spatial (marked) point processes. In this section, we give some definitions and  notations (mainly from Ilian et al. \cite{IllianEtAl}) for spatial unmarked and marked point processes and introduce some summary functions.
 Furthermore, we recall some thinning operations for point processes. For more rigorous treatment of the topic, the reader is referred to Illian et al.\  \cite{IllianEtAl}, Diggle  \cite{diggle_book}, M{\o}ller and Waagepetersen  \cite{moller}, and Chiu et al.\  \cite{chiu2013stochastic}. 
 
\subsection{\em Spatial point processes} 

\noindent Spatial point processes are mathematical models for point patterns. A spatial point process $X$ is defined as the random set of locations in a spatial domain $D$ where  the events of interest occur. Usually, the point process is observed in an observation window $W\subset D$. We refer to this as a point pattern in the observation window $W$. In this work, the locations, where the nerves enter the epidermis and where they terminate, are treated as realisations of spatial point processes in a rectangular observation window $W\subset\R^2$. The point processes are assumed to be {\em locally finite}, that is for every bounded subset  $B$ in the Borel set $\BB(\R^2)$, the number of points of the process that lie in $B$ is finite.
Further, the point processes are assumed to be $simple$, that is at any location there is at most one point of the process. Lastly, the point processes are assumed to be stationary (translation invariant) and isotropic (rotation invariant). 

Sometimes additional characteristics, marks, are attached to each point in the point pattern. Such marked point patterns often provide a deeper insight into the underlying physiological processes \cite{IllianEtAl}. 
Let $M \subset \mathbb{R}$ be a mark space with the mark $m_{i}\in M$ attached to the point $x_{i} \in X$.
A realisation of the corresponding marked point process $X_M$ is then 
\begin{equation*}
\{(x_{i},m_{i}), i=1,...,n\} \subset W \times M,
\end{equation*}
where $n$ is the observed number of points. We assume that $X_M$ is stationary and isotropic, i.e.\ invariant under translations and rotations of the point locations, respectively.

\subsection{\em Ripley's K function}\label{summary}

Here, we review Ripley's $K$ function that is used to describe the second-order properties of a point  process \cite{ripley76}. For stationary and isotropic point processes in $\R^2$, Ripley's $K$ function has a straightforward interpretation. In particular, $\lambda K(r)$, $\lambda$ being the intensity (the mean number of points per unit area) of the process, gives the expected number of further points of the process in the disc with radius $r$ centred at an arbitrary point of the process.  For the Poisson point process, 
\begin{equation*}
    K(r) = \pi r^2 ,\qquad r\geq 0.
\end{equation*}
 The homogeneous Poisson point process corresponds to complete spatial randomness, and therefore it is often treated as a reference model \cite{IllianEtAl,miles1970homogeneous}.  Observed values  that are smaller or larger than their theoretical values under complete spatial randomness indicate regularity or clustering, respectively. Since there are points of the process outside the observation window that may interact with the points inside the window,  an edge correction term is needed when estimating the $K$ function. An approximately unbiased estimator of the $K$ function is given by 
\begin{equation*}
    \hat{K}(r)=\frac{\mid W\mid }{ n^2} \sum_{i=1}^n\sum_{j\neq i}w(x_i,x_j) \1\{\mid\mid x_i-x_j \mid\mid\leq r\},\quad r\geq 0,
\end{equation*}
where $n$ is the total number of the observed points in $W$, $||x_i-x_j||$ denotes the Euclidean distance between the points $x_i$ and $x_j$, $\1\{A\}$ is the indicator function equal to 1 when event $A$ is true and zero otherwise, and $w(x_i,x_j)$ is an edge correction term. We used the translation correction $w(x_i,x_j)=\frac{1}{\mid W_{x_i}\cap W_{x_j}\mid}$, where $W_{x_i}$ is the translated window $W_{x_i}=\{z+x_i: z\in W\}$ and $\mid\cdot\mid$ denotes the two dimensional Lebesgue measure.  In this work, we use the variance stabilized and centred variant of the $K$ function \cite{IllianEtAl}, defined by
\begin{equation}
 \label{eq:L}
    L(r)-r = \sqrt{\frac{K(r)}{\pi}}-r, \quad r\geq 0, 
\end{equation}
which for  the Poisson process equals zero. Therefore, positive values of this centred function indicate clustering and negative values regularity.

Our data are hierarchically structured into groups (healthy, mild, moderate), subjects within the groups, and samples from the subjects. Since we are interested in the average spatial structure of the ENFs in each group, the overall group-wise $K$ and $L$ functions need to be estimated.  This can be achieved as follows. For group $g$, we initially estimate the sample-wise summary functions $K_{ij}$ for sample $j\in\{1,...,m_i\}$ of subject $i$, $i\in\{1,...,N_g\}$. Let $\hat{K}_{ij}$ denote the corresponding estimator.

 Then, the subject specific mean function $\Bar{K_i}$ can be obtained as a weighted mean of the functions $K_{ij}$. An unbiased estimator for the subject-wise $\Bar{K}_i$ function for each subject $i$ is given by 
\begin{equation*}
    \widehat{\bar{K}}_i(r)= \sum_{j=1}^{m_i}w_{ij}\hat{K}_{ij}(r).
\end{equation*}
Similarly, the subject-wise functions are weighted to obtain the group-wise function $\Bar{K}_g$ for the group $g$. A group-wise estimator for $\Bar{K}_g$ is given by 
\begin{equation}
\label{eq:groupsum}
    \widehat{\bar{K}}_g(r) = \sum_{i=1}^{N_g}w_{i}\hat{K}_{i}(r).
\end{equation}
Square point number weights are used to compute the subject-wise and group-wise  estimates, since the point patterns from different samples and subjects cannot be assumed to have the same intensity \cite{diggle2000comparison,myllymaki2012analysis,konstantinouspatial}. Let $n_{ij}$ denote the number of points in sample $j$ of subject $i$, and let $n_i=\sum_{j=1}^{m_i} n _{ij}$ be the total number of points in the samples from subject $i$. Then, the square point number weights for the group-wise  and subject-wise $K$ functions are given by 
\begin{equation*}
    w_i =\frac{n_i^2}{\sum_{k=1}^N n_k^2},
    \quad w_{ij}=\frac{n_{ij}^2}{\sum_{k=1}^{m_i}n_{ik}^2}.
\end{equation*}
Second-order properties of point processes with marks can also be investigated.
The mark correlation function for marked point processes is discussed in the following section. 

\subsection{Mark correlation function}

The mark correlation function describes the second-order characteristics for point processes with quantitative marks. It can be defined as the (conditional) expectation 
 \begin{equation*}\label{mark correlation}
k(r)=\frac{\E [f(m_{i},m_{j}) | \| x_{i}-x_{j}\| =r]}{\mu ^{2}},
\end{equation*}
where $\mu$ is the mean of the considered mark distribution and $ f(m_{i},m_{j})$ is a so-called test function \cite{StoyanStoyanRatio}. 
Here, we use $f(m_{i},m_{j})=m_{i} m_{j}$. 
Therefore, if the marks are uncorrelated, $k$ equals 1. Values less than 1 indicate negative correlation and values greater than 1 positive correlation. 
The mark correlation function can be estimated by kernel estimation, namely 
\begin{equation}
\label{eq:markcor}
\hat{k}(r)=\frac{\sum\limits_{i=1}^n \sum\limits_{j=1,j\neq i}^{n}m_{i}m_{j} \cdot w_{ij}}{\bar m^2},
\end{equation}
where $\bar m$ is the mean of the observed marks,
\begin{equation*}
w_{ij}=\frac{e_{b}(r-|x_{i}-x_{j}|)}{|W_{i} \cap W_{j}|},
\end{equation*}
and $e_{b}$ is the Epanecnikov kernel function with bandwidth $b$ \cite{StoyanStoyanRatio}. The bandwidth can be chosen e.g.\ by using the rule-of-thumb given in \cite{silverman1986}.

Since our data are replicated, we need to estimate the mark correlation function for each sample and then pool all the estimates to obtain the subject-wise and group-wise estimates in a similar fashion as we estimate the $K$ function above. The mark correlation ( Eq. \eqref{eq:markcor}) and the $L$ function (Eq. \eqref{eq:L}) will be used throughout the paper to assess the goodness of fit of the proposed model.
The inference method we chose, on the other hand, requires an informative summary function of the data.
To avoid using the same summary functions for inference and model evaluation, we used a different summary function for inference, which is described in the following section.

\subsection{Empty space function}
The \textit{empty space distribution function} $F(r):[0,\infty)\rightarrow[0,1]$ is related to the probability that an arbitrary point $x\in \mathbb R^2$ has an empty disc of radius $r$ around it.  It is  defined as
\begin{equation*}
    F(r) = 1- P(X(b(x,r))=0),
\end{equation*}
where $X(b(x,r))$ is the random number of points of the process in the disc centered at $x$ with radius $r$, denoted by $b(x,r)$. Let $\{y_i\}_{i=1}^t$ be $t$  points randomly sampled within $W_{\ominus r}:=\{x\in W : min(\mid\mid x-x_b\mid\mid)\geq r,\quad x_b\in \partial W\}$, where $\partial W$ is the boundary of $W$.  Then, an unbiased estimator  for the empty space function $F$ \cite{IllianEtAl} is given by

\begin{equation}
    \hat{F}(r) = \frac{1}{t}\sum_{i=1}^{t} \1\{min(\mid\mid y_i- x\mid\mid)\leq r, \,\, x\in X\cap W\}.
    \label{eq:F_hat}
\end{equation}

\section{Modelling the ENF thinning process}
Thinned point processes provide a class of models for point patterns that are caused by random mortality.  A thinning operation defines a rule which determines which points in a point process $X$ should be deleted, to obtain a thinned point process $X_{thin}$.  Thinning operations can be divided into the following three different types \cite{IllianEtAl}: 
\begin{itemize}
    \item {\em \textbf{Independent p-thinning:}}
In $p$-thinning each point in the point process $X$ is deleted with constant probability $1-p$, $p\in[0,1]$, independently of its location and on the other points in $X$. The parameter $p$ is called the ``retention probability''.
\item {\em \textbf{Independent $\pi(x)$-thinning:}}
A generalisation of the $p$-thinning  is the $\pi(x)$-thinning. In $\pi(x)$-thinning the retention probability depends on the location $x$ of the point, that is for all $x\in X$, the deterministic function $\pi(x)$ gives the probability that  $x\in X_{thin}$. As $p$-thinning, $\pi(x)$ thinning is  statistically independent, that is the deletion or non-deletion of any particular point does not depend on the operation on the other points.
\item {\em \textbf{Dependent thinning:}}
More general thinning strategies can be constructed if we let the retention probability to depend on the other   points in the point process $X$, that is for every point $x\in X$ the retention probability is given by a function $\pi(x\mid X)$.
\end{itemize}

As neuropathy advances, the number of nerve trees (base points) and end points decreases. Here, we suggest  a thinning scheme to describe the biological process behind these changes. It is believed that whole ENF trees die and, in addition, some individual nerve endings may  disappear or appear, the latter being caused by the existing nerve fibers branching and creating new end points in order to compensate for the loss of nerves. In addition, the spatial pattern of base and end points becomes more clustered as the neuropathy advances as illustrated in Figure \ref{fig:Lfun}.  Both the base and end point patterns are clustered as their corresponding centered $L$ functions ( Equation \eqref{eq:L} ) are positive, except at very small distances.  Note that the clustering of end points increases from healthy to mild and from mild to moderate patterns but the clustering of base points increases only from healthy to mild.

Below, we first illustrate that  an independent random thinning scheme, e.g.\ an independent p-thinning or $\pi(x)$-thinning, applied to healthy samples does not result in patterns similar to the mild patterns.  Then, we propose a dependent thinning strategy,  where the probability for a point to be retained depends on the other points in $X$, and suggest an approximate Bayesian computation approach for the inference of the model.

\subsection{Independent random $p$-thinning}

A natural starting point is to investigate whether mild patterns can be constructed by randomly removing either complete nerve trees or individual nerve end points from healthy patterns,  i.e.\ that there is no underlying mechanism that guides the nerve removal. However, if the process $X$ is stationary, then Ripley's $K$ and $L$ functions are invariant under the independent random thinning operation, and therefore summary functions of $X$ and of the thinned process $X_{thin}$ are identical. As mild patterns of base and end points are more clustered than the healthy patterns, we expect  the independent random thinning to be unsuitable for  capturing the spatial structure of the mild diabetic patterns. 

\begin{figure}[H]
    \centering
  \subfloat{\includegraphics[scale=0.5
  ]{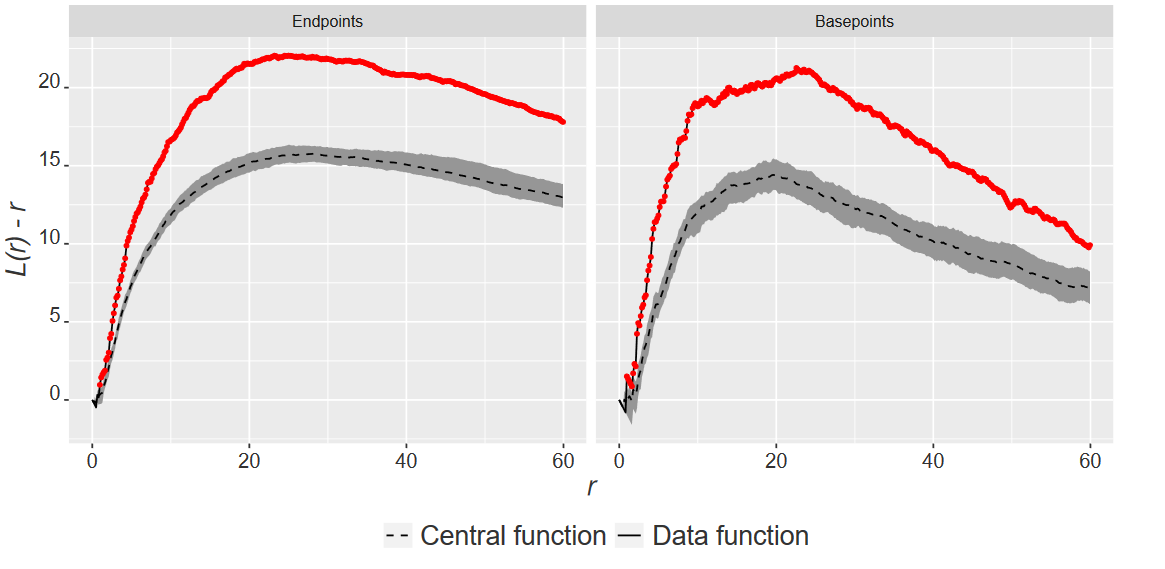}}
\caption{ Group-wise centred $L$ functions based on the data (red, solid) for the mild  patterns with $95\%$ global envelopes (grey) based on independent random p-thinning of  end points (left)  or base points (right) of the healthy patterns.}
    \label{fig:independent_thinning}
\end{figure}

To confirm this, we chose as our null models two different independent random $p$-thinning models and applied them to the healthy base and end point patterns. For each model this was performed by (i) estimating the probability $p$ as the ratio between the corresponding mean mild group and mean healthy group intensities $\Bar{\lambda}_M$ and $\Bar{\lambda}_H$, and (ii) by randomly and independently removing either end points or base points together with all the connected end points with probability $1-p$. Then, we constructed $95\%$ global envelopes (see Appendix) based on 2500 independent summary curves obtained via simulations from an independent thinning model with the estimated retention probability $\hat{p}$ (see Figure \ref{fig:independent_thinning}).  Since the data curves (in red) fall completely outside the envelopes, we can conclude that, as expected, the thinned normal patterns fail to capture the structure present in the mild patterns.

\subsection{Dependent thinning}
\label{section:dependent_thinning}
\begin{figure}[H]
    \centering
    \includegraphics[scale=0.65]{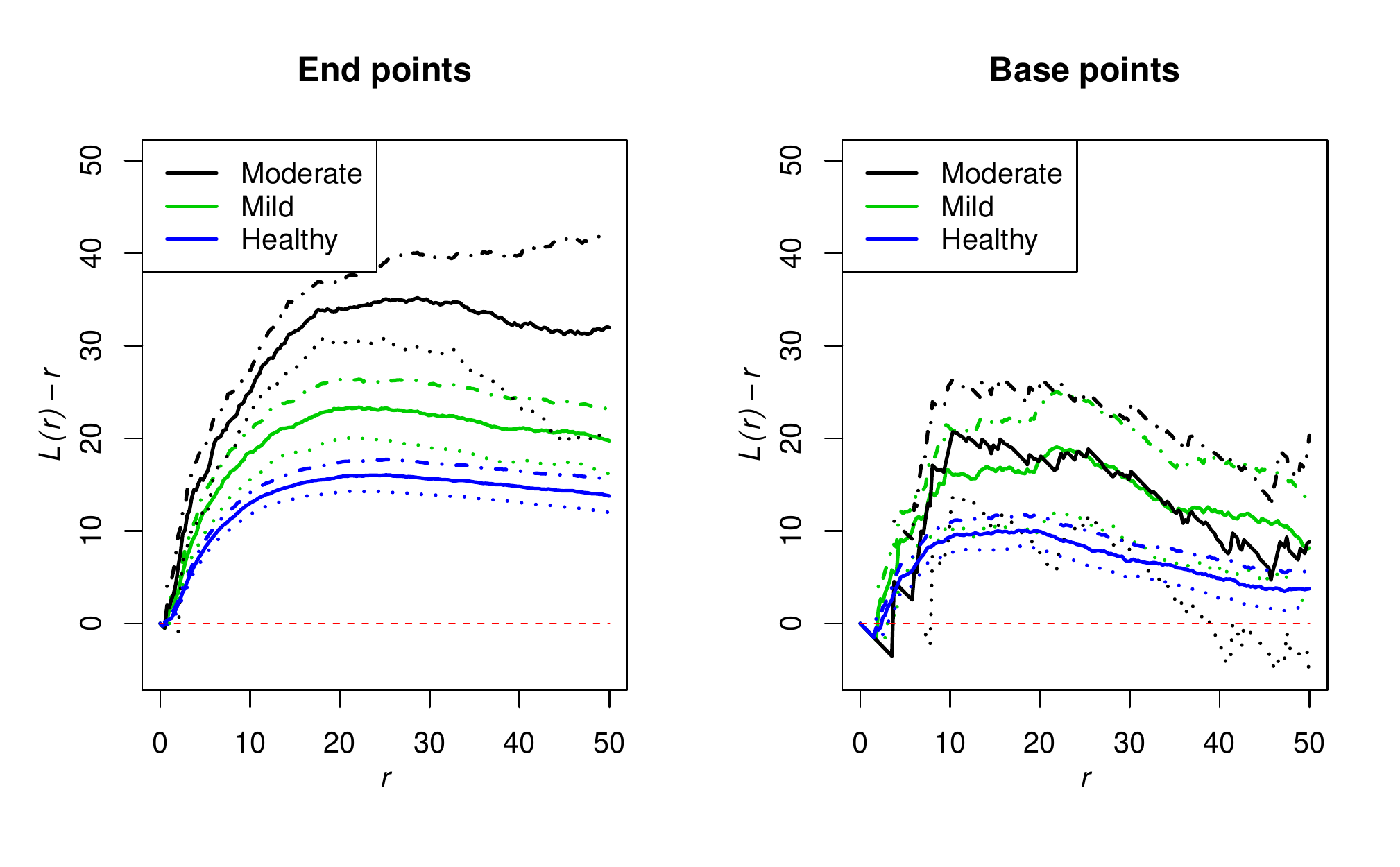}
    \caption{Group-wise pooled $L(r)-r$ functions, for the end points (left) and base points (right) of the healthy (blue), mild (green) and moderate (black) groups with 95$\%$ pointwise bootstrap envelopes (dashed lines). The red dashed line corresponds to the complete spatial randomness.}
    \label{fig:Lfun}
\end{figure}

It can be seen in Figure \ref{fig:Lfun} that the mild base and end point patterns are more clustered than the corresponding healthy patterns. Therefore, base/end points should be removed from healthy patterns such that the resulting patterns are more clustered. Below, we suggest a parametric thinning strategy, where isolated nerve trees are more likely to be removed than non-isolated ones.  

 Let $B^{M}$ and $E^{M}$ denote the base and end point patterns for the targeted mild diabetic sample observed in $W$ and $n_{B}$ and $n_{E}$ be the numbers of points in $B^{M}$ and $E^{M}$, respectively. Now, let $B^{H}=\{y_{j}, m(y_{j})\}$ and $E^{H} = \{x_{i}, m(x_{i})\}$, with $j=1,\dots,n_B'$ and $i=1,\dots,n_E'$, denote the marked base and end point patterns for a healthy sample, with mark $m(y_{j})$ giving the Euclidean distance to the closest other base point $y_{k}$, $k\neq j$, with $n_{B}'$ and $n_{E}'$ being the numbers of points in $B^{H}$ and $E^{H}$, respectively.

  We thin the pattern $B^{H}$ to exactly $n_{B}$ base points according to an iterative thinning scheme (see Algorithm \ref{alg:model}), with $\pi(y_{j}) = f(m(y_{j});\theta)$, where $\theta$ is a scale parameter and $f(\cdot;\theta)$ is given by 
\begin{equation}
    f(m;\theta)\propto e^{-\theta^2m^2}.
    \label{EQ:probability}
\end{equation} 

\begin{algorithm}
\caption{Dependent thinning model $\mathcal{M}(\theta\mid B^H,n_B)$}\label{alg:model}
\begin{algorithmic}
     \State \textbf{Input}: Healthy basepoint pattern $B^H=\{y,m(y)\}$ and
    corresponding endpoint pattern $E^H=\{x,m(x)\}$, parameter $\theta$, desired number of basepoints $n_B$.
\State \textbf{Output}: Simulated mild diabetic basepoint pattern $y^*$ and corresponding endpoint pattern $x^*$.

\Repeat \\
\qquad Let $I =\{1,\dots,\mid y \mid\}$ be an index set.\\
    \qquad For $y_j\in y$, $j\in I$, \ calculate \ $m_j=m(y_j)>0$\\
     \qquad   For $y_j\in y$, $j\in I$,\ calculate  $f_j=\frac{1-e^{-\theta^2 m_j^2}}{\sum_k 1- e^{-\theta^2 m_k^2}}$ 
     \\
     \qquad Sample with replacement an index $l\in I$ using the weights $f_j$, $j\in I$. \\
     \qquad Remove $y_l$ from $y$. Call $y^*$ the resulting basepoint pattern.  \\
     \qquad  From the endpoint pattern $x$, remove all points that are connected to $y_l$. Call $x^*$ the \\\qquad resulting endpoint pattern.
     \\ \qquad Update the basepoint pattern $y=y^*$ and the endpoint pattern $x=x^*$
        \Until $| B^H| = n_B$\\
  \end{algorithmic}
\end{algorithm}

\begin{figure}[H]
    \centering
    \includegraphics[scale=0.5]{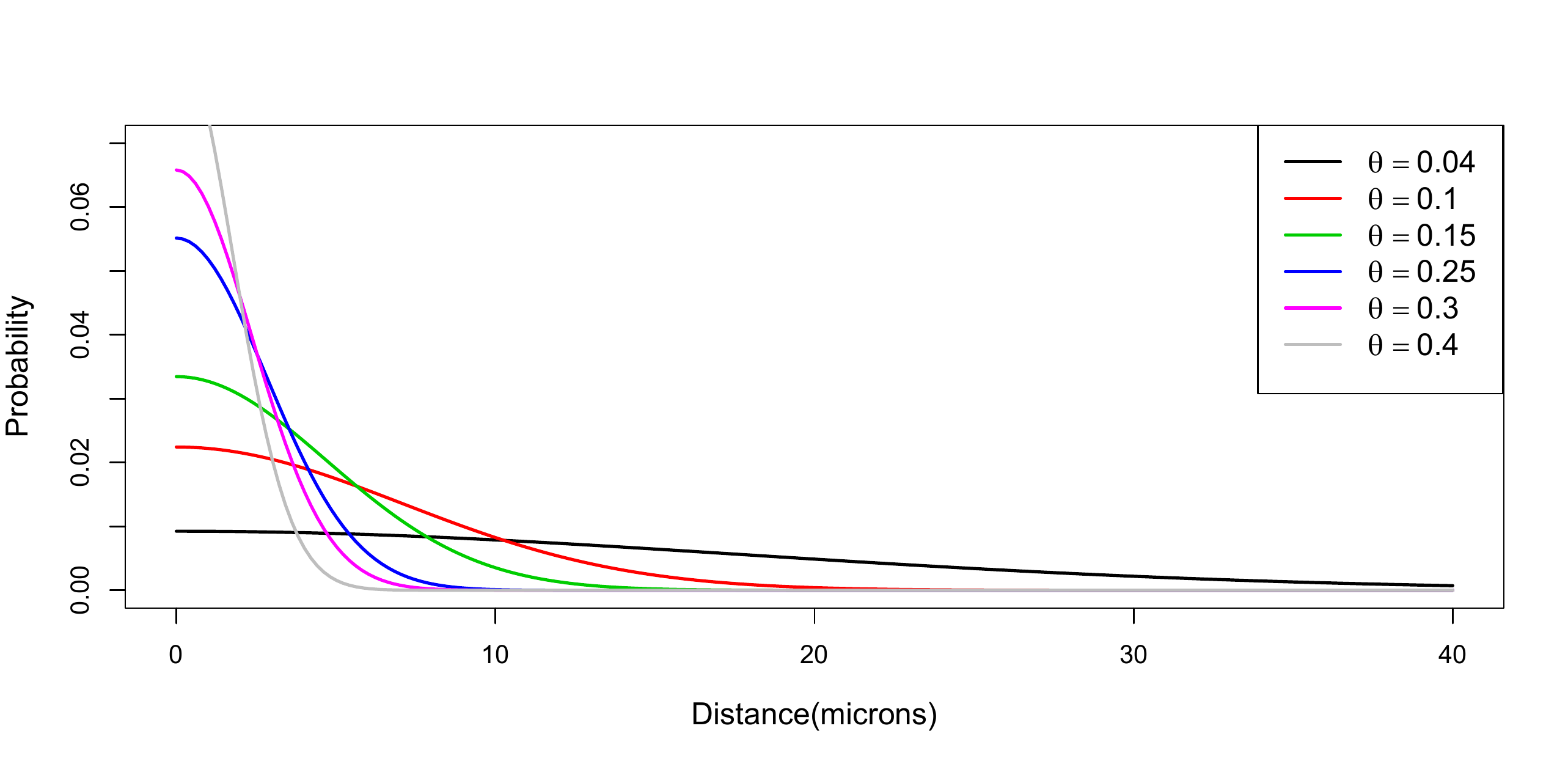}
    \caption{The density function defined in Equation \eqref{EQ:probability} for different values of the parameter $\theta$.}
    \label{fig:Probab}
\end{figure}

 By definition, $f(\cdot;\theta)$ favors the removal of isolated points, i.e.\ points with large marks $m$, since the retention probabilities $\pi(y_j)$ decrease with increasing distance $m$ (see Figure \ref{fig:Probab}). Moreover, the removal probabilities are proportional to $1-e^{-\theta^2m^2}$, and therefore, the larger the value of $\theta$, the closer this thinning strategy is to independent thinning . For each $y_{j}$ that is removed, we remove all the end points connected to it. The resulting base and end point patterns are denoted by $y^*$ and $x^*$, respectively.

For a mild diabetic sample with $n_B$ base points, the aforementioned thinning model is applied to all healthy patterns that have at least $n_B+5$ base points. The number of such healthy patterns for a mild pattern $a\in\{1,...,28\}$ is denoted by  $N_a$. Hence for each mild diabetic sample $a$, $N_a$ thinned replicates with exactly $n_B$ base points are constructed.  Then, these $\prod_{a=1}^{28} N_a$  thinnings are used to construct group-wise $L$ function estimates. Throughout the remainder of this paper, this model will be denoted as $\mathcal{M}(\theta) = \mathcal{M}(\theta\mid B^H,n_B)$, with parameter $\theta$, and is detailed in Algorithm \ref{alg:model}.

\subsection{ Inference using approximate Bayesian computation}

To infer plausible values of the scale parameter $\theta$ controlling the retention probabilities, we used approximate Bayesian computation (ABC) \cite{sisson2018handbook,marin2012approximate}. This is a family of algorithms suitable for Bayesian inference when the likelihood function associated to a statistical model $\mathcal{M}(\theta)$ is unavailable in closed form, or is computationally too expensive to approximate,  but given a parameter vector $\theta$, it is possible to simulate artificial data from $\mathcal{M}(\theta)$. The simplest ABC method is the acceptance-rejection sampling \cite{pritchard1999population}, that, for given data $y$ and (vector of) summary statistics thereof $S(y)$, samples from an approximation $P_\epsilon(\theta\mid S(y))$ of the posterior distribution $P(\theta\mid S(y))$. This is performed by (i) proposing a parameter $\theta^*\sim P(\theta)$ sampled from its prior distribution $P(\theta)$ ; (ii) conditionally on $\theta^*$, simulate an artificial dataset $y^*$ as $\mathcal{M}(\theta^*)\rightarrow y^*$, to be read as the output of a ``run'' of model $\mathcal{M}(\theta^*)$; (iii)  reduce both $y^*$ and $y$  to a low-dimensional set of summary statistics $S(y^*)$ and $S(y)$, respectively, and evaluate their proximity  using some distance (e.g. Euclidean) $\| S(y^*)-S(y)\|$;  and finally, (iv) retaine $\theta^*$ if $\| S(y^*)-S(y)\| < \epsilon$, for some small $\epsilon>0$, and reject it otherwise. The procedure (i)--(iv) is iterated until $N$ parameter values have been accepted. A pseudocode for the ABC rejection sampler is given in Algorithm \ref{alg:cap} below. Each accepted parameter is a draw from the approximate posterior
\begin{equation*}
    P_\epsilon(\theta\mid S(y)) \propto P(\theta)\int  \1_{\mid\mid s^* - s)\mid\mid < \epsilon} P(s^*|\theta)ds^*,
\end{equation*} 
where in the integrand we have used the shorthand notations $s^*=S(y^*)$ and $s=S(y)$. This algorithm is computationally inefficient when the posterior is very dissimilar to the prior, resulting in many rejections. Instead of fixing $\epsilon$, we can simulate a large number of $\theta$ values and choose the most appropriate of these as described below.

More sophisticated ABC methods taking into account information about the previously accepted draws for $\theta$ have been suggested, both in an MCMC framework \cite{marjoram2003markov,picchini2014inference} and as sequential Monte Carlo algorithms (\cite{sisson2007sequential}, \cite{beaumont2009adaptive}, \cite{del2012adaptive}). However, as our model includes only the scaling parameter $\theta$, the simple ABC rejection based method described above was found to be suitable enough.

\begin{algorithm}
\caption{ABC rejection sampler}\label{alg:cap}
\begin{algorithmic}
\State \textbf{Input}: prior $P(\theta)$, model $\mathcal{M}(\theta)$, summaries $S(\cdot)$, threshold $\varepsilon>0$, positive integer $N$.
\State \textbf{Output}: posterior draws $(\theta_1,...,\theta_N)$.
      \For{\texttt{$i \gets 1,...,N$}}
        \Repeat \\
    \qquad Draw from prior $\theta^* \sim P(\theta)$\\
     \qquad   Simulate  $\mathcal{M}(\theta^*) \to y^*$\\
     \qquad Compute $S(y^*)$
        \Until $\| S(y^*)-S(y)\| < \epsilon$\\
       $ \theta_i \gets \theta^*$
      \EndFor
\end{algorithmic}
\end{algorithm}
Notice that $P_\epsilon(\theta\mid S(y))$ coincides with $P_\epsilon(\theta\mid y)$, when $S(\cdot)$ is a sufficient statistic for $\theta$. On the other hand, sufficient statistics are generally unavailable and therefore, in practice, ABC always returns approximate inference even in the limit when $\epsilon=0$. It is therefore crucial to construct appropriate (``informative'') summary statistics that are able to retain information about $\theta$. As a rule of thumb, it is suggested that the length of the vector $S(y)$ (which is the same as the length of $S(y^*)$) should be the same as the length  of $\theta$  \cite{fearnhead2012constructing}. 

In our case, as we only have the scaling  parameter $\theta$ in Equation  \eqref{EQ:probability}, we construct a single summary statistic.  The $K$ or $L$ function could be chosen as the summary statistic in the ABC algorithm but since the centred $L$ function will be used to evaluate the goodness-of-fit of the thinning model, we chose to use the empty space function $F_y$ instead.  However, we did not use the entire function but we considered the summary function previously used in \cite{kuronen2021}. In particular, we used $s=S(y)=\min( \{ r : \hat{F}_y(r) = 0.3\})$, where $\hat{F}_y(r)$ is the estimator given in Equation  \eqref{eq:F_hat}. The summary statistic for the observed data is $s=S(y)=\min( \{ r : \hat{F}_y(r) = 0.3\})$, where $\hat{F}_y(r)$ is the estimator given in Equation  \eqref{eq:F_hat}, and $y$ are the observed mild diabetic basepoint patterns. 
Similarly, for generic simulated mild base point patterns $y^*$ obtained using $\mathcal{M}(\theta^*\mid B^H,n_B)$, as defined in Section \ref{section:dependent_thinning}, we computed the summary statistics $s^*=S(y^*)=\min( \{ r : \hat{F}_{y^*}(r)=0.3\})$. Notice that, even though $\mathcal{M}(\theta^*\mid B^H,n_B)$ generates a simulated endpoint pattern $x^*$, the inference is solely based on the simulated and empirical mild basepoint patterns $y^*$ and $y$.

\subsection{ABC inference using simulated data}

 A simulation study was conducted to assess the performance of the inference method. For this purpose, healthy data were simulated from a Mat\'ern cluster process using parameters estimated from the data using the minimum contrast method \cite{IllianEtAl}. The simulated parent pattern, was represented by the base points, and the simulated daughter pattern, was represented by the end points.  Then, the proposed dependent thinning was applied to the simulated pattern for different chosen values for $\theta$, and the corresponding realisations were used to obtain the empirical summary statistics used in the ABC method.
 
\begin{figure}[H]
    \centering
  \subfloat{\includegraphics{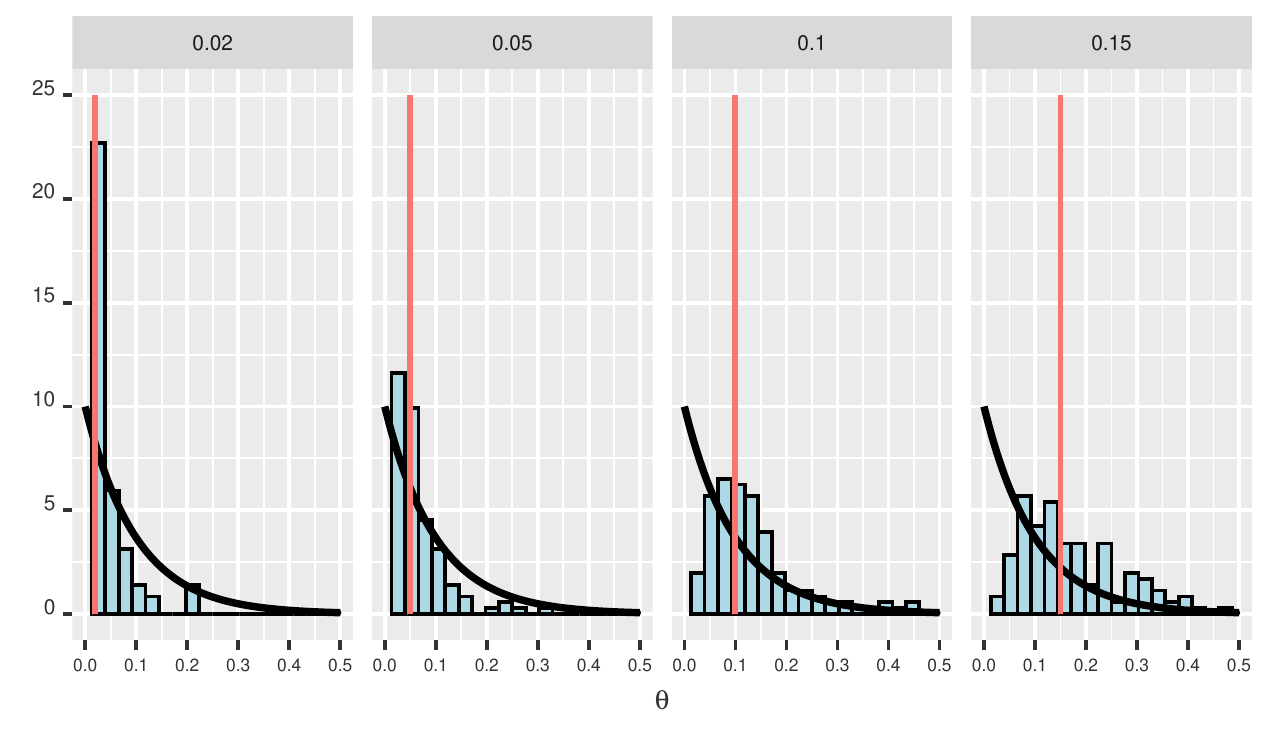}}\qquad
    \caption{Histograms of ABC posterior draws for $\theta$ from a simulation study. The prior density (solid black line) is an Exponential(10) truncated to values larger than $0.01$ and the true values of $\theta$ (solid red lines) are also reported on the panel headings. }
   \label{fig:Sim_study_ABC}
\end{figure}

 Each pattern was thinned such that $n_B=14$ parent points (and their daughter points) remained. The posterior distributions for $\theta$ are displayed in Figure \ref{fig:Sim_study_ABC} together with an exponential prior $P(\theta)=\mathrm{Exponential}(10)$ for $\theta$.  The true values of the data-generating $\theta$ are  included for comparison. We observe that the true parameter value $\theta^o$ is well identified when $\theta^o<0.1$, and while for larger values the posterior mode is close to the true value, the posterior uncertainty increases, that is as the value of $\theta^o$ increases it becomes progressively more challenging to identify it. This is due to the fact that as $\theta^o$ increases the dependent thinning scheme  approaches independent random thinning, as shown in Figure \ref{fig:Probab}, and as a result, the influence of the parameter on the realisations diminishes.
Inference results have been obtained using the ``reference table'' version of the ABC rejection algorithm, which does not require the threshold $\varepsilon$ to be prefixed in advance, and it is illustrated as follows:
(i) we simulated $1,330,000$ parameters independently from the prior Exponential(10), and conditionally on these draws we simulated correspondingly $1,330,000$ data sets $y^*=\mathcal{M}(\theta^*)$; (ii) we calculated $S(y^*)$ on each of these data sets and $S(y)$ for the observed data, and (iii) accepted those $\theta^*$'s for which the corresponding distances $\| S(y)-S(y^*)\|$ were smaller than the $0.1$-th percentile of all $1,330,000$ ABC distances. We used the \texttt{abc} function in the R \texttt{abc} package \cite{abc-package} to carry out the computations. 
In next section we consider inference on real ENF data.

\begin{table}[H]
\centering
\caption{Posterior median and $95\%$ credibility intervals  $CI_{\theta}$ for the parameter $\theta$ in the simulation study, where the target patterns have 14 base points and the true value of $\theta$ varies between 0.02 and 0.15. }
\begin{tabular}{c|c|c }

 True value of  $\theta$ & Median & $CI_{\theta}$ \\
 \hline
  0.02 & 0.028 & [ 0.011, 0.195 ]  \\
  0.05 & 0.038& [ 0.011, 0.211 ] \\
 0.10 &  0.119& [ 0.037, 0.384 ] \\
  0.15 & 0.150& [ 0.046, 0.404 ]
\end{tabular}

\label{table:par}
\end{table}

\section{Results}
In this section, the dependent thinning model is fitted to the ENF data and the goodness-of-fit of the model is investigated by comparing  the thinned and target patterns with respect to some spatial and non-spatial summary statistics. Initially, we compare the structure in the thinned healthy and mild patterns, and then, in the  thinned mild and moderate patterns. 

\subsection{\em Healthy vs Mild}

\begin{figure}[H]
    \centering
  \includegraphics[scale=1]{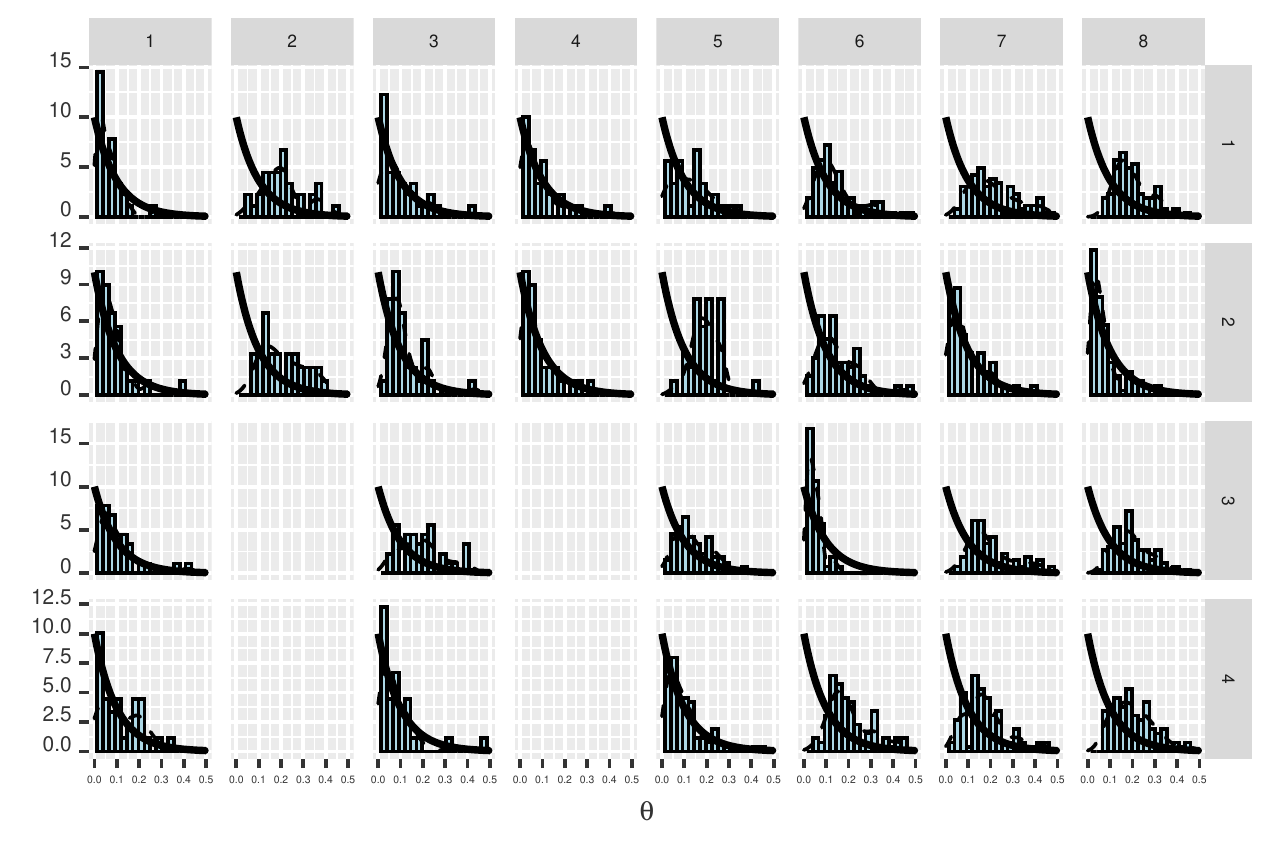}
     \caption{ENF data: histograms of ABC posterior draws for $\theta$  for each mild diabetic sample obtained by thinning healthy patterns. The prior densities are plotted with solid black lines. Each column corresponds to one subject (total 8 subjects), and each row corresponds to different samples from one subject. The number of samples for each subject are varying from 2 to 4.}
    \label{fig:params}
\end{figure}

We applied the thinning strategy introduced in Section \ref{section:dependent_thinning} and removed whole nerve trees (base points and end points connected to it) from the healthy patterns.
The posterior distributions of $\theta$ parameter based on the thinned healthy patterns are displayed for each mild diabetic pattern in Figure \ref{fig:params}.  As the model favors the removal of isolated trees if the $\theta$ is small, and  corresponds to independent thinning when $\theta$ is large, an Exponential(10) prior  was chosen for $\theta$. Initially, we used a uniform prior which gave  posterior distributions with large variance. We believe that this choice made the ABC method more efficient, which resulted in better inference. For several mild diabetic samples, the bulk of the posteriors $P_\epsilon(\theta\mid S(y))$ is located around very small $\theta$ values indicating that, on a typical healthy pattern, isolated nerve trees are favored to be removed from the healthy patterns in order to obtain patterns similar to these mild patterns. On the other hand, for some mild diabetic samples, the posterior $P_\epsilon(\theta\mid S(y))$ is centred at "large" values, indicating that randomly thinning the healthy patterns is sufficient to capture the structure in the targeted mild diabetic sample. The latter patterns contained rather large number of nerve trees indicating that the neuropathy is in an early stage, and hence cannot be detected from the nerve patterns yet. Also for a few mild diabetic samples, the posterior $P_\epsilon(\theta\mid S(y))$ coincides with the prior. In particular, the inference quality is low for the patterns with small number of nerve trees or for the patterns where most of the nerve trees are located close to the edge of the observation window.

\begin{figure}[H]
    \centering
\subfloat{\includegraphics{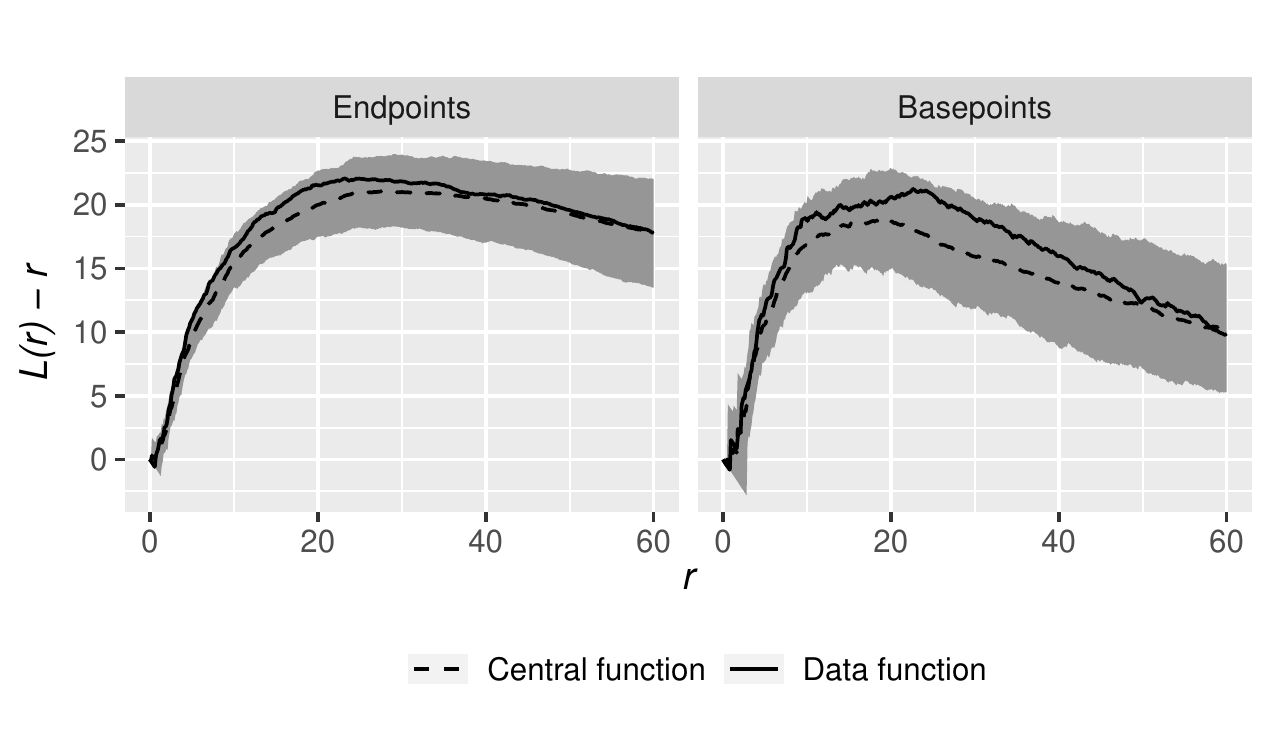}}\qquad
    \caption{Group-wise centered $L$ functions with $95\%$ global envelopes for the end points (left) and base points (right) constructed from 2500 simulations from the posterior predictive distribution of the thinning model. The solid curves are the centered $L$ functions estimated from the mild data.}
    \label{fig:group_thin}
\end{figure}

The results regarding  the spatial structure of the end and base points of the thinned healthy patterns are presented in Figure \ref{fig:group_thin}.  The proposed thinning scheme creates  patterns that capture both the end point and base point spatial structure very well as the $95\%$ global envelopes completely cover the empirical centered $L$ curves. The envelopes are based on 2,500 simulations from the posterior predictive distribution of the thinning model. In other words, for each mild diabetic neuropathy sample, we simulate data $y^*$ using $\mathcal{\theta^*}$, with $\theta^*$ sampled from $P_{\epsilon}(\theta\mid S(y))$, which are then used to calculate the  mean $L(r)-r$ for the simulated mild diabetic neuropathy group.  Notice that each of the 2500 simulations is generated by selecting one of the approximately 1300 posterior draws available for each mild diabetic sample and one of the 112 healthy diabetic samples at random.  A pseudocode for this procedure is given in Algorithm \ref{alg:bands}.

\begin{figure}[H]
    \centering
  \subfloat{\includegraphics[scale=0.8]{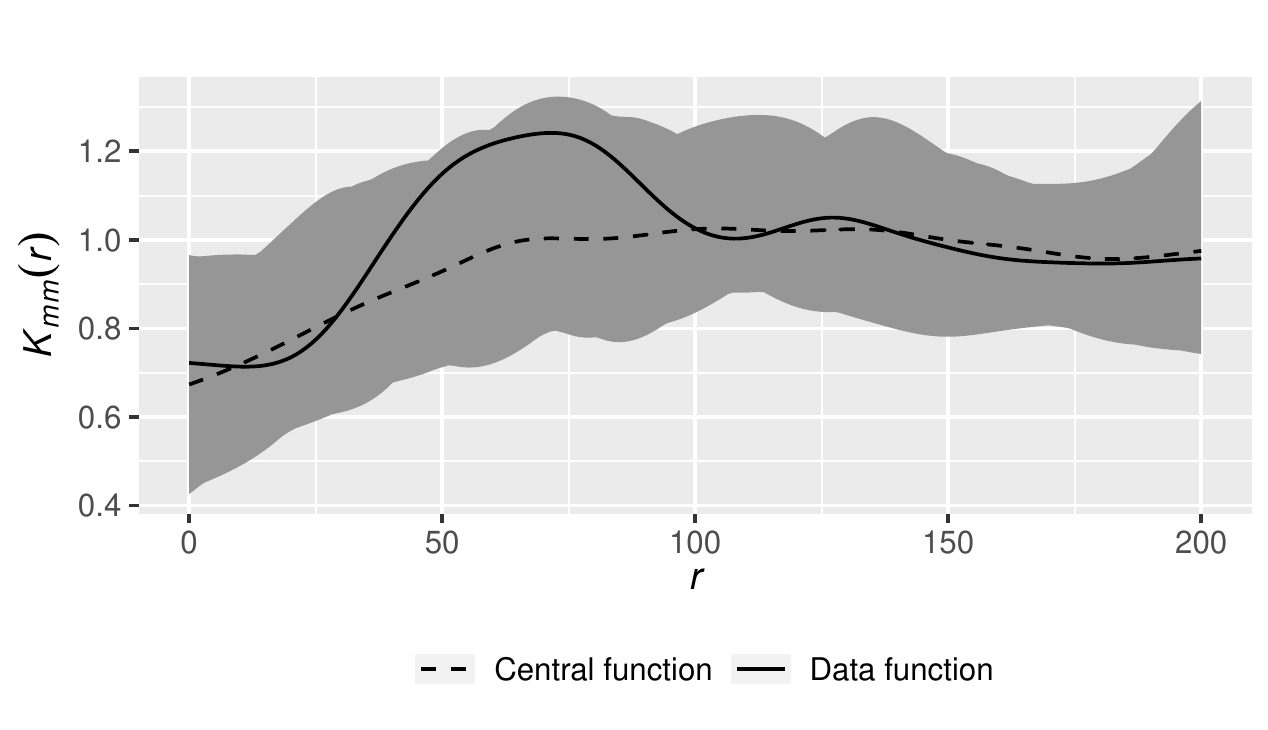}}\qquad
    \caption{Group-wise mark correlation function for the base points marked by the areas of their reactive territories  with $95\%$ global envelopes constructed from 2500 simulations from the posterior predictive distribution of the thinning model. The solid curve is the mark correlation function estimated from the mild data.}
    
    \label{fig:markcorr}
\end{figure}

\begin{algorithm}
\caption{Posterior predictive bands}\label{alg:bands}
\begin{algorithmic}
\State \textbf{Input}: Basepoint and endpoint healthy patterns $\{B^H,E^H\}_j$ with $j=1,\dots,112$, \\ Number of basepoints $n_B^i$ in the mild diabetic samples $i=1,\dots,28$,\\
 Posterior sample $\theta^i=(\theta_1^i,\dots,\theta_N^i)$ for the mild diabetic samples $i=1,\dots,28$, \\
 Summary function $T(\cdot)$.
\State \textbf{Output}: $95\%$ global envelopes constructed from 2500 simulations from the posterior predictive distribution.
      \For{\texttt{$s \gets 1,...,2500$}}
       \For{\texttt{$i \gets 1,...,28$}} \\
       \qquad Sample with replacement a value $\theta$ from $\theta^i$\\
    \qquad Sample with replacement an index $j\in\{1,\dots,112\}$\\
     \qquad   Simulate
       $\mathcal{M}(\theta\mid B^H_j,n_B^i)\to (y^*, x^*)$\\
       \qquad Compute $T_i=T(y^*,x^*)$
      \EndFor \\
      Calculate $\Bar{T}_g^s$ using $(T_1,\dots,T_{28})$ as described in Equation \eqref{eq:groupsum}
      \EndFor \\
      Construct $95\%$ global envelopes (Appendix \ref{glenvtest}) using  $(\Bar{T}_g^1,\dots, \Bar{T}_g^{2500})$
\end{algorithmic}
\end{algorithm}
Moreover, for each nerve tree we calculated the area of its reactive territory - convex hull determined by the base and its end points - which was then attached to each base point as a mark. To deal with zero areas, i.e.\ nerve trees with only one end point, the length between the end and the base point was used instead. This length is much smaller than a typical area of a reactive territory. Figure \ref{fig:markcorr} illustrates the mark correlation function with $95\%$ global envelopes constructed using simulations from the posterior predictive distribution as explained above. The thinning model captures even the mark correlation structure between the sizes of the reactive territories well.


\begin{figure}[H]
    \centering
    \includegraphics[scale=0.8]{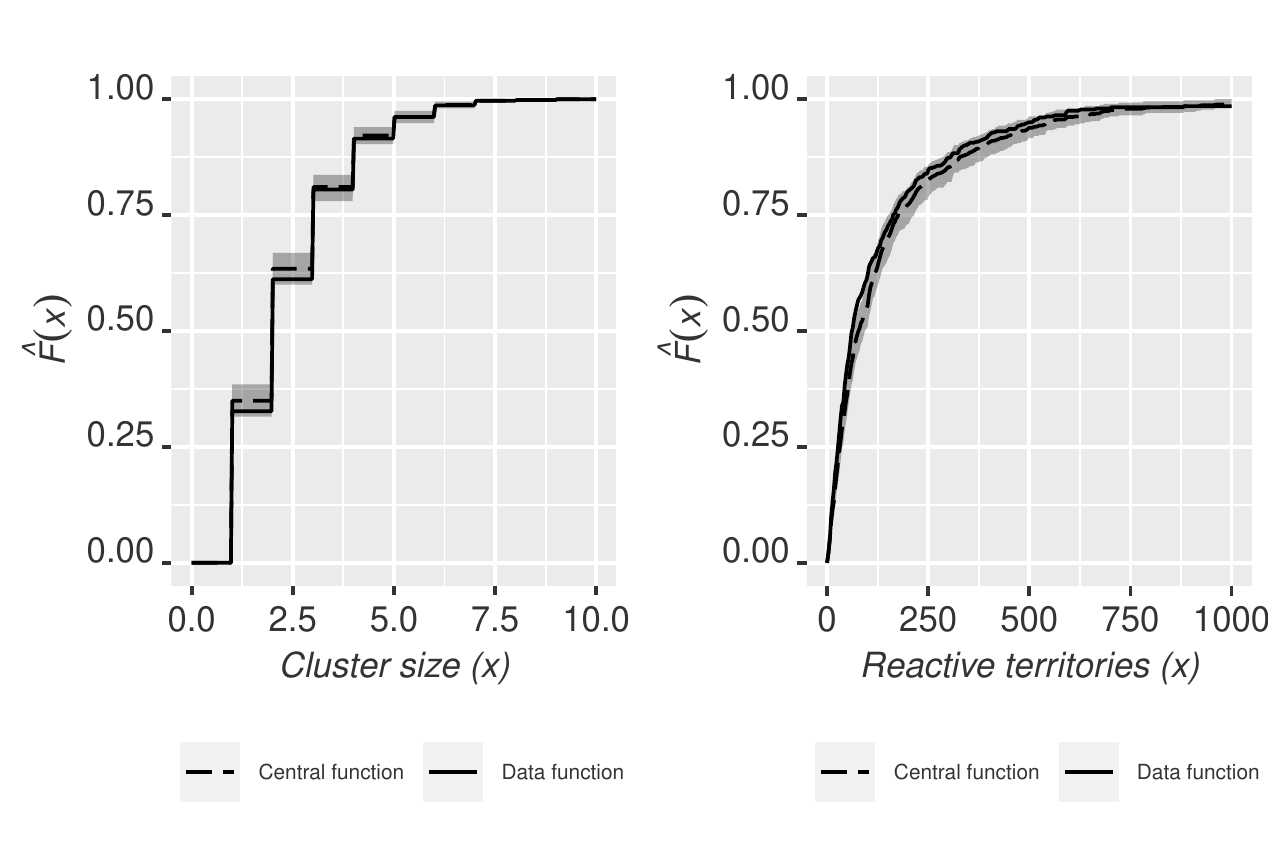}
   \caption{ Cumulative distribution functions of the cluster size (left) and the total  area of reactive territories (right)  with $95\%$ global envelopes constructed from 2500 simulations from the posterior predictive distribution of the model. The solid curves are the corresponding cumulative distribution functions estimated from the mild data.}
    \label{fig:cluster_area_thinned_normal}
\end{figure}

We also computed some non-spatial summary statistics, namely  the cluster size distribution, i.e.\ the cumulative distribution of the number of end points per nerve tree, and of the total area of the reactive territories to evaluate the dependent thinning model,   
 illustrated in Figure \ref{fig:cluster_area_thinned_normal}. The model seems to capture even these characteristics well.

\subsection{\em Mild vs Moderate}
When applied to the healthy patterns, the suggested dependent thinning approach seems to be able to produce patterns similar to the empirical mild patterns. A natural question is whether we can obtain patterns similar to the observed moderate patterns by thinning mild patterns in a similar manner. It can be seen in Figure \ref{fig:Lfun} that the end points in the moderate patterns are more clustered than in the mild patterns but since there is not such a big difference in clustering of the base points in the two groups, independent random thinning could be appropriate for the base points. 
 Therefore, we randomly thinned the mild patterns by removing nerve trees to the number of nerve trees, i.e.\ base points, in the observed moderate patterns. Note that only the base points, not the end points, are randomly thinned. No parameters need to be estimated in this case as nerve trees are randomly thinned with equal probabilities.

\begin{figure}[H]
    \centering
  \subfloat{\includegraphics[scale=0.6]{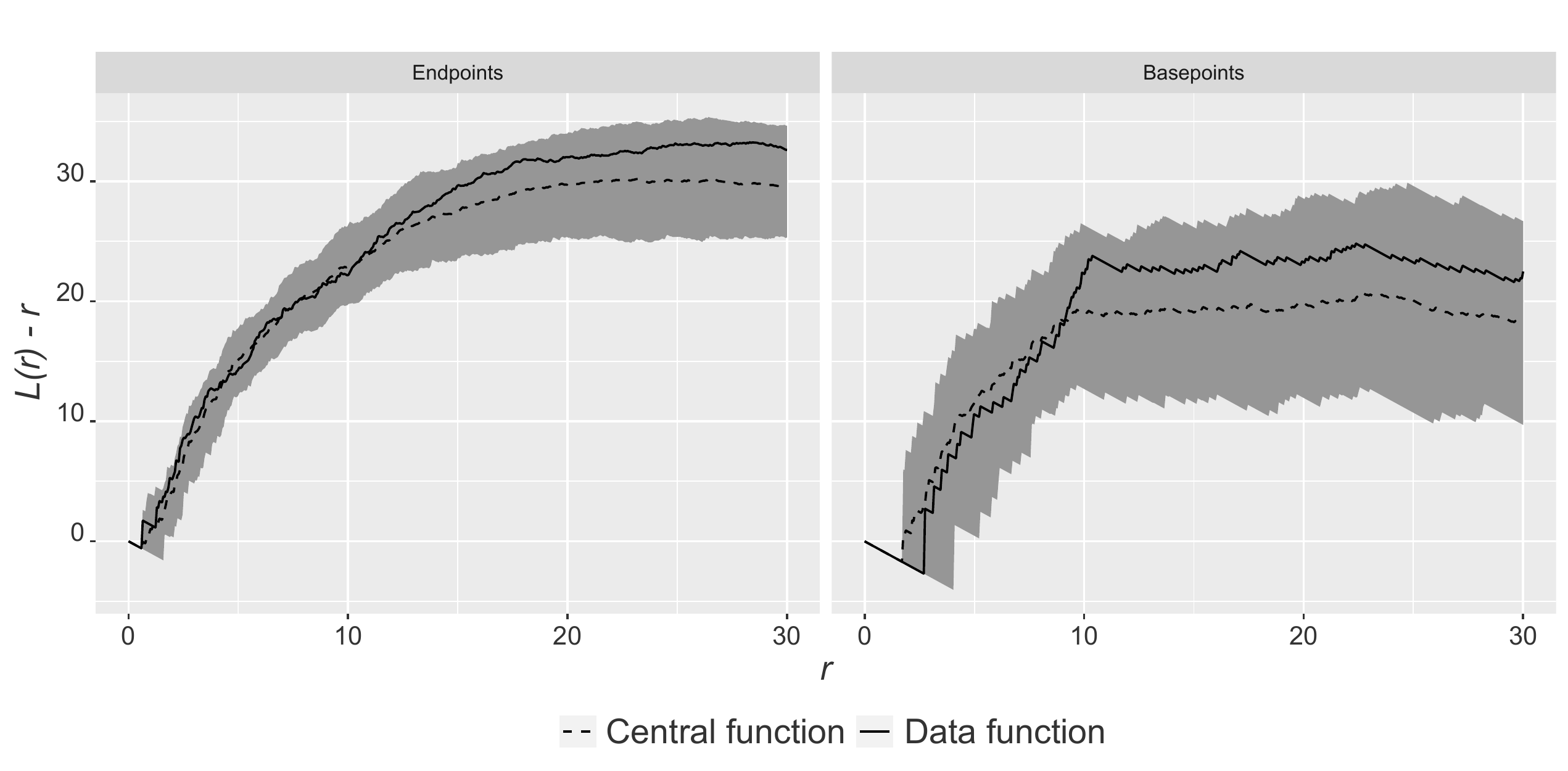}}

    \caption{Group-wise centered $L$ functions with $95\%$ global envelopes for the end points (left) and base points (right) constructed from 2500 simulations from the independent thinning model. The solid curves are the centered $L$ functions estimated from the moderate data. }
    \label{fig:independent_thinning_num}
\end{figure}

\begin{figure}[H]
    \centering
  \subfloat{\includegraphics[scale=0.5]{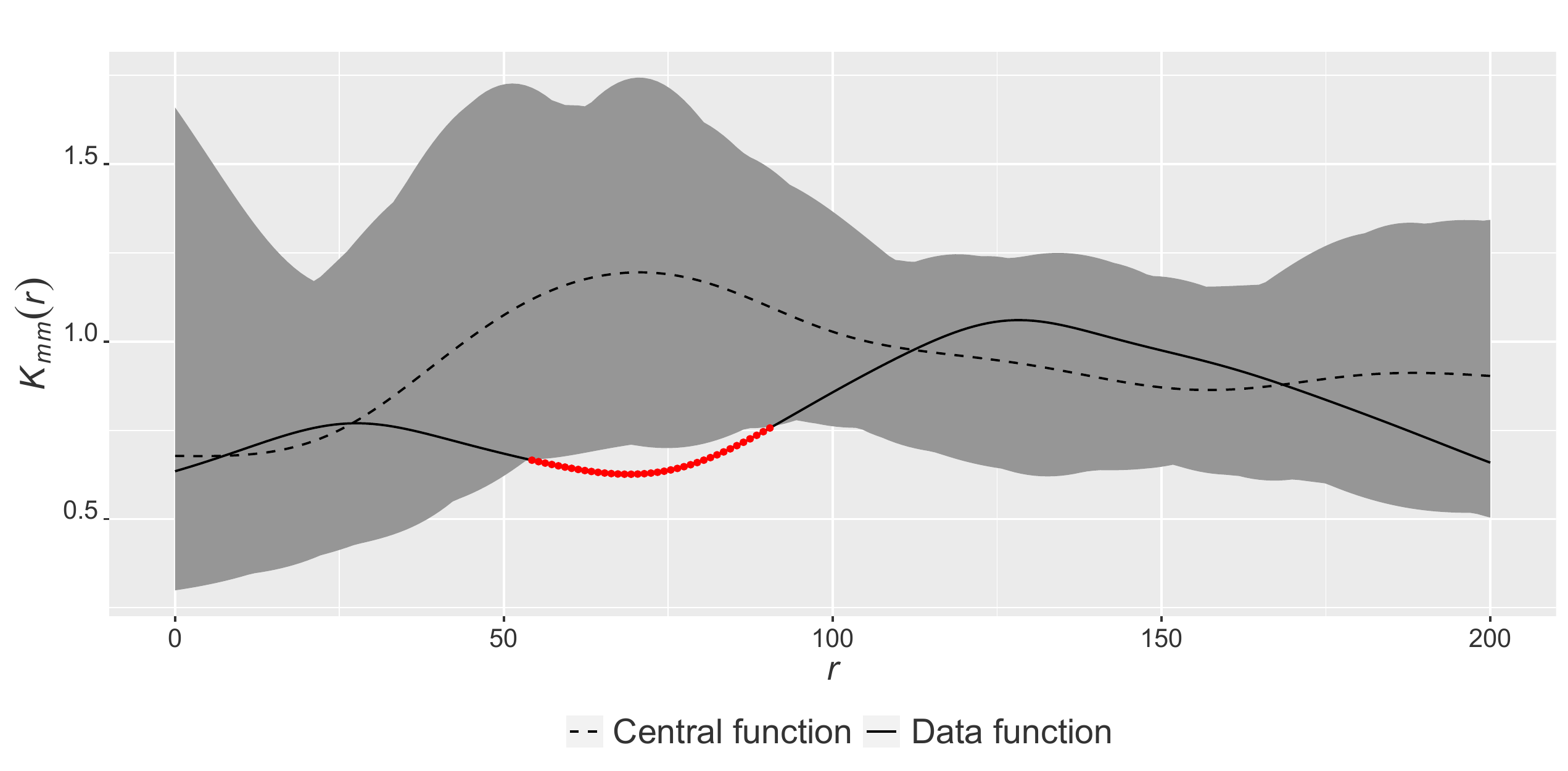}}\qquad
    \caption{Group-wise mark correlation function for the base points marked by the areas of their reactive territories  with $95\%$ global envelopes constructed from 2500 simulations from the independent random thinning model. The solid curve is the mark correlation function estimated from the moderate data.}
    
    \label{fig:markcorr_moderate}
\end{figure}

\begin{figure}[H]
    \centering
  \includegraphics[scale=0.29]{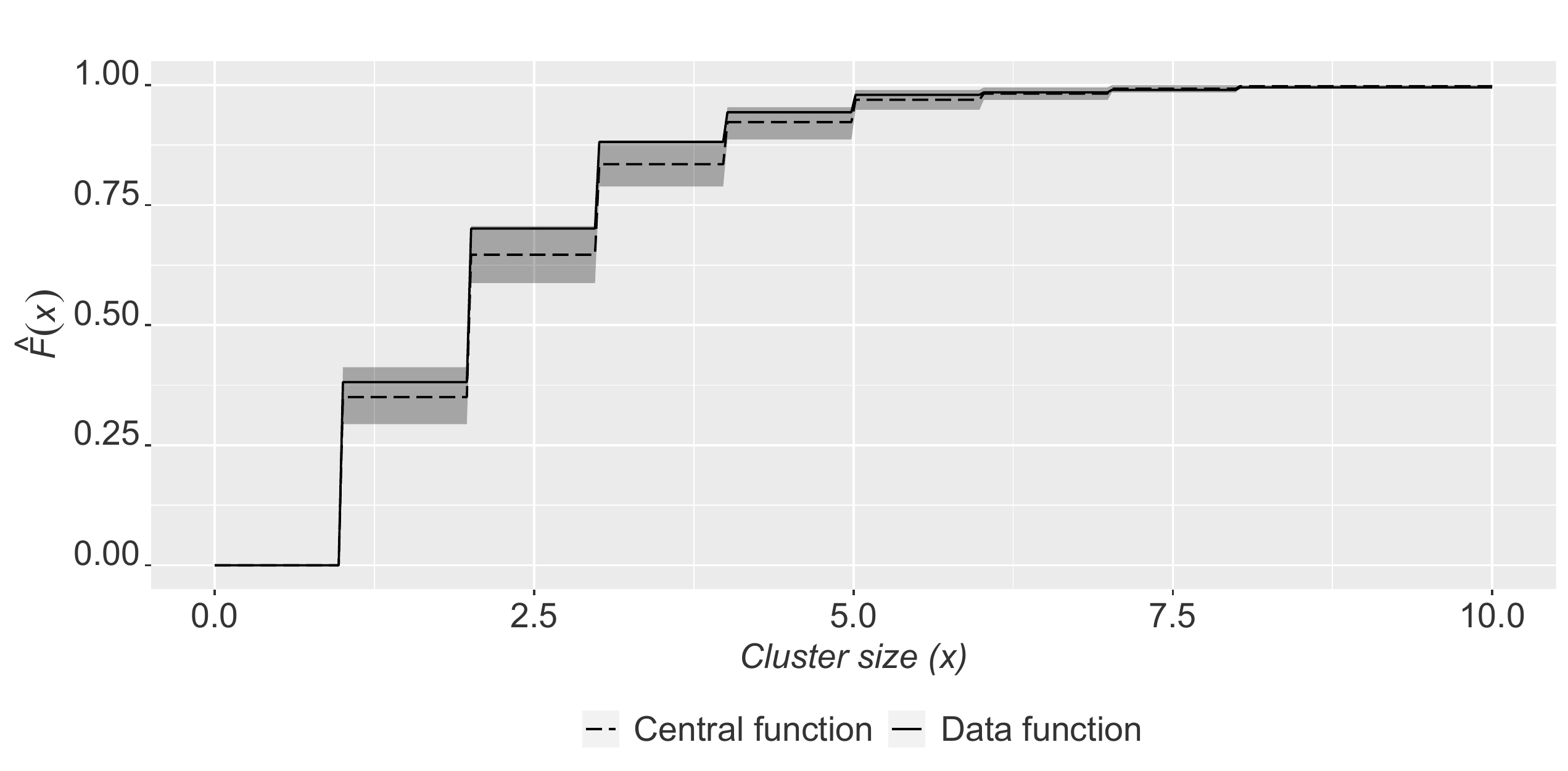}
 \subfloat{\includegraphics[scale=0.29]{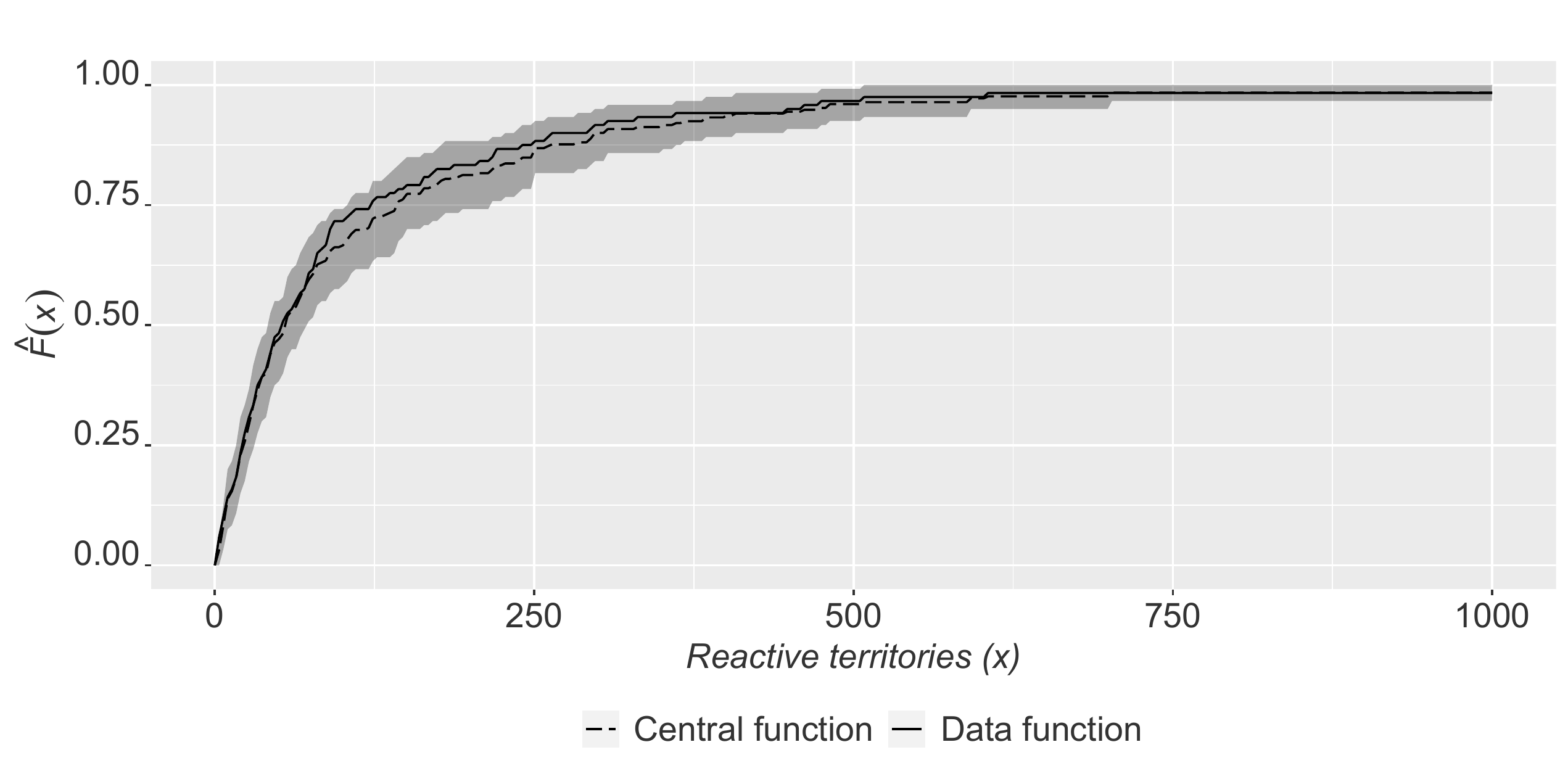}}
    \caption{Cumulative distribution functions of the cluster size (left) and the total  area of reactive territories (right)  with $95\%$ global envelopes constructed from 2500 simulations from the independent random thinning model. The solid curves are the corresponding cumulative distribution functions estimated from the moderate data.}
    \label{fig:cluster_area_moderate}
\end{figure}

The group-wise centered $L$ functions with $95\%$ global envelopes for the base and end point patterns after independent thinning of nerve trees are given in Figure \ref{fig:independent_thinning_num}. We observe that the empirical centered $L$ functions lie within the envelopes for the base point patterns indicating a good fit of the model. Even the overall structure of the end point patterns is captured quite well by the model. However, the mark correlation of the sizes of the reactive territories is not completely caught by the independent thinning model, see Figure \ref{fig:markcorr_moderate}. On the other hand, as seen in Figure \ref{fig:cluster_area_moderate}, the end point cluster size distribution and the distribution of the area of the reactive territories are quite well described by the model even though in the former case the moderate data based distribution is very close to the upper boundary of the interval.

\section{Discussion}
The biological process that guides the physiological changes in the epidermal nerve fiber structure of the neuropathy was investigated. The effects of varying severity of the underlying neuropathic condition, diabetes, were also analyzed. For this purpose, we treated the ENF samples from the feet of healthy patients, and from patients with mild and moderate diabetic neuropathy, as realisations of spatial point processes occurring in response to progressive pathologic severity. As the spatial intensity of the ENFs decreases with progression of neuropathy, the mild diabetic neuropathy patterns can be considered as spatial thinnings of the healthy patterns, and the moderate diabetic neuropathy patterns as spatial thinnings of the mild  patterns. Therefore, we proposed spatial thinning models for the changes that occur in the ENF structure in neuropathy as the as diabetes becomes more severe. To the best of our knowledge, this is the first study that considers such spatial thinning models to investigate the nerve removal as a result of varying severity of diabetes, and corresponding severity of neuropathy.

Two spatial thinning models were investigated, an independent random $p$-thinning and a dependent thinning scheme. The scale parameter controlling the retention probability in the latter model was estimated by using approximate Bayesian computation (ABC), which is a very flexible methodology that is applicable whenever the likelihood function is unavailable or is computationally expensive to evaluate but it is feasible to simulate from a computer model.

We  focused first on the nerve removal from the healthy patterns to obtain patterns similar to the observed mild patterns. An independent random $p$-thinning seems insufficient in modeling the nerve removal process at this earliest stage of diabetic neuropathy. Therefore, a more complex thinning model that favored the removal of isolated nerves in order to increase the overall clustering in the patterns, was proposed.  Measured by the $L$ function, mark correlation function, and some non-spatial summary statistics, the model was able to describe the change from healthy to mild diabetic neuropathy very well. On the other hand, the independent removal of nerve trees was enough to model the nerve mortality in mild diabetic neuropathy patients to obtain patterns similar to the observed moderate patterns. 

Our original hypothesis was that first, whole ENF trees die due to the neuropathy and then, some individual nerve endings may disappear or appear. In our study, it was enough to remove entire nerve trees and no additional removal or addition of individual nerve endings was needed. However, we can see in Figure \ref{fig:cluster_area_moderate} that the independent thinning of mild patterns only barely covered the corresponding curve estimated from the moderate patterns and in Figure \ref{fig:markcorr_moderate} that the fit of the mark correlation function was not perfect in this case. Therefore, even though the suggested models describe the data sufficiently well, they could still be improved.

\section*{Acknowledgements}
The authors thank William R.\ Kennedy's group (University of Minnesota) for blister
immunostaining, quantification and morphometry of the ENF data. 
The authors also thank the Swedish Research Council for financially supporting the project. UP acknowledges funding from the Swedish National Research Council (Vetenskapsrådet 2019-03924) and the Chalmers AI Research Centre.
\section*{Data availability statement}
Unfortunately, we are not able to share the data publicly.

\newcommand{\SortNoop}[1]{}

\bibliographystyle{plain}

\appendix

\section{Appendix: Examples of ENF samples}\label{enfsamples}
\begin{figure}[h]
    \centering\includegraphics[scale=0.8]{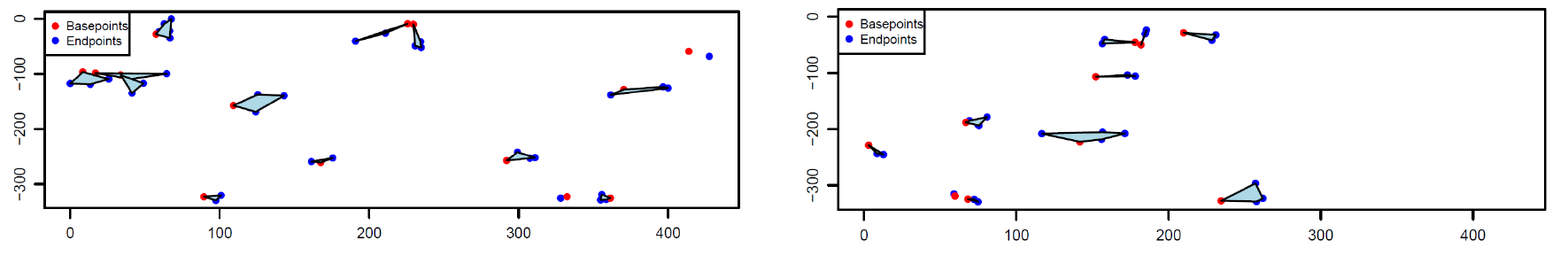}\caption{Examples of ENF patterns for mild (left) and moderate (right) diabetic samples.}
    \label{fig:samples_ex}
\end{figure}

\section{Appendix: Global envelope tests}
\label{glenvtest}

Global envelope tests are graphical Monte-Carlo tests based on functional or multivariate summary statistics \cite{Myllymaki}. Let $T_1, T_2,...,T_s$ be $d$-dimensional vectors containing the discretizations of a functional statistic, such as the $K$ or $L$ function, at $d$ points, with $T_1$ denoting the data vector and $T_2,...,T_s$ the vectors obtained under the model corresponding to a null hypothesis (``null model''). An envelope is defined as the band bounded from below by the $d$-dimensional vector $T_{low}$ and  from above by the $d$-dimensional vector $T_{upp}$. Then, a  $100(1-\alpha)\%$ global envelope is a set ($T_{low}^\alpha,T_{upp}^\alpha)$ such that the probability that $T_i$ falls outside the envelope in any of the $d$ points is $\alpha$.  In the recent point process literature, such tests are widely used to assess the goodness-of-fit of spatial point process models, as well as constructing global confidence bands from a set of functions obtained from the posterior predictive distribution. Construction of a global envelope  ($T_{low}^\alpha,T_{upp}^\alpha)$ depends on the measure used to rank the extremeness of the vectors $T_1, T_2,...,T_s$ as well as on the significance level $\alpha$. 
Here, we used the extreme rank length measure (ERL) to construct $95\%$ global envelopes \cite{Myllymaki,narisetty2016extremal}.  Using the ERL measure the different statistics are initially ranked for each value $r$ in the discretization grid. Then, the number of $r$ values where the statistic is extreme is taken into account to construct the ERL measure $E_i$ for each statistic $T_1, T_2,...,T_s$, with $E_i< E_j$ interpreted as $T_i$ is more extreme than $T_j$. 
Now let $E_{(\alpha)}\in\mathbb{R}$ be the largest $E_i$ such that
\begin{equation*}
    \sum_{i=1}^s \1(E_i<E_{(\alpha)})\leq \alpha s
    \label{eq:ea}
\end{equation*} 
and let  $I_{(\alpha)}$ be the set of vectors less than or as extreme as $E_{(\alpha)}$. Then, the $100(1-\alpha)\%$ global envelope based on the ERL measure is given by \begin{equation*}
    \left(T^{(\alpha)}_{low \ k},T^{(\alpha)}_{upp \ k}\right) =
\left(\min_{i\in I_{(\alpha)}}T_{ik},\max_{i\in I_{(\alpha)}}T_{ik}\right)\quad \text{for $k=1,...,d$}.
\label{envelope}
\end{equation*}

\end{document}